\documentclass[pra,showpacs]{revtex4}
\usepackage{amssymb}
\usepackage[dvips]{graphicx}
\usepackage{amsmath}
\usepackage{graphicx}
\usepackage{makeidx}
\usepackage{subfigure}

\begin{document}

\title{Two-dimensional dissipative gap solitons}
\author{Hidetsugu Sakaguchi$^{1}$ and Boris A. Malomed$^{2}$}
\affiliation{$^{1}$Department of Applied Science for Electronics and Materials,
Interdisciplinary Graduate School of Engineering Sciences, Kyushu
University, Kasuga, Fukuoka 816-8580, Japan\\
$^{2}$Department of Physical Electronics, School of Electrical Engineering,
Faculty of Engineering, Tel Aviv University, Tel Aviv 69978, Israel}

\begin{abstract}
We introduce a model which integrates the complex Ginzburg-Landau
(CGL) equation in two dimensions (2D) with the linear-cubic-quintic
combination of loss and gain terms, self-defocusing nonlinearity,
and a periodic potential. In this system, stable 2D
\textit{dissipative gap solitons} (DGSs) are constructed, both
fundamental and vortical ones. The soliton families belong to the
first finite bandgap of the system's linear spectrum. The solutions
are obtained in a numerical form and also by means of an analytical
approximation, which combines the variational description of the
shape of the fundamental and vortical solitons and the balance
equation for their total power. The analytical results agree with
numerical findings. The model may be implemented as a laser medium
in a bulk self-defocusing optical waveguide equipped with a
transverse 2D grating, the predicted DGSs representing spatial
solitons in this setting.
\end{abstract}

\maketitle

\section{Introduction and the model}

Equations of the complex Ginzburg-Landau (CGL) type are universal asymptotic
models to describe the nonlinear pattern formation in dissipative media \cite%
{reviews2}. They also find direct (rather than asymptotically derived)
realizations in nonlinear optics as models of laser cavities \cite{reviews1}%
. Objects of fundamental interest predicted by the CGL equations are
solitary pulses (SPs), alias dissipative solitons. They represent, in
particular, temporal pulses generated by fiber lasers \cite%
{reviews1,laser,Kutz} and, in an altogether different physical context,
patches of traveling-wave thermal convection in \ narrow channels \cite%
{Kolodner}.

The simplest CGL equation is based on the cubic nonlinearity. Exact
one-dimensional (1D) SP solutions are known in that case
\cite{Lennart}, but they are unstable, as the respective equation
includes the linear gain, compensating the cubic loss, which
destabilizes the zero background around the pulses and leads to the
formation of a chaotic ``gas" of
SPs \cite{Jena}. A well-known modified equation that can support \emph{stable%
} SPs includes a combination of linear and quintic loss terms and cubic
gain, therefore it is called the cubic-quintic (CQ) CGL equation. This
equation was introduced (in the 2D form) in Ref. \cite{NizhnyNovgorod}.
Stable SP solutions to the 1D version of the CQ-CGL equation were first
predicted by means of an analytical approximation based on the power-balance
analysis for solitons of the nonlinear Schr\"{o}dinger (NLS) equation \cite%
{me}. Then, these solutions were explored in detail by means of numerical
methods \cite{CQCGL}. More general models, such as linearly coupled systems
of CQ-CGL equations \cite{Ariel} and the complex Swift-Hohenberg equation
with the CQ nonlinearity \cite{SakaguchiBrandt}, were introduced too.

In addition to the solitons of the NLS type, a generic species of solitary
waves in conservative media is represented by gap solitons. They are well
known in nonlinear optics, as temporal solitons in fiber Bragg gratings, and
in Bose-Einstein condensates (BECs) with repulsion between atoms, which are
loaded into in optical-lattice trapping potentials. A fundamental feature,
from which the name of the gap soliton derives, is that its wavenumber (in
terms of optical models) must belong to a finite bandgap induced by the
effective periodic potential. In the context of both the fiber gratings \cite%
{FBG1,FBG2,GapModel} and matter waves (BEC) \cite{BEC,Sa1}, gap solitons
were predicted theoretically, including 2D gap solitons in BEC \cite%
{Salerno,BEC-2D}, and 2D solitons with embedded vorticity \cite{vortex}. The
creation of optical and matter-wave gap solitons was reported in the 1D
geometry -- in short fiber gratings \cite{exper-FBG}, and in the condensate
loaded into an optical lattice combined with a strong transverse trap \cite%
{Markus}, respectively.

A generalization of the above concepts, aiming to predict \textit{%
dissipative gap solitons} (DGSs), was proposed recently \cite{we}. The
respective version of the CGL equation combines a periodic potential and the
set of the CQ loss and gain terms:%
\begin{eqnarray}
i\frac{\partial \psi }{\partial t} &=&-\frac{1}{2}\frac{\partial ^{2}\psi }{%
\partial x^{2}}+|\psi |^{2}\psi -A\cos \left[ 2q_{0}\left( x-\frac{L}{2}%
\right) \right] \psi  \notag \\
&&+i\epsilon (-\gamma _{1}+\gamma _{2}|\psi |^{2}-\gamma _{3}|\psi
|^{4})\psi .  \label{1D}
\end{eqnarray}%
Here, $\psi $ may be considered as the local amplitude of the
electromagnetic wave (in the case of an optical model), $A$ and $\pi /q_{0}$
are the strength and period of the potential ($x=L/2$ is the midpoint of the
system), and coefficients $\epsilon \gamma _{1},$ $\epsilon \gamma _{3}$ and
$\epsilon \gamma _{2}$ represent the linear and quintic dissipation and
cubic gain, respectively. The conservative cubic term in Eq. (\ref{1D}) is
defined with the self-defocusing sign, which is relevant to gap solitons.

A physical realization of Eq. (\ref{1D}) pertains to a planar
self-defocusing optical waveguide, in which the periodic potential
may be induced by a transverse grating (periodic modulation of the
refractive index). In particular, a self-defocusing nonlinearity, in
the combination with photoinduced transverse lattices, can be
implemented in photorefractive crystals \cite{photo}. As concerns
the CQ loss/gain terms, they may actually represent a combination of
the linear amplification and saturable absorption, which is a common
setting in laser cavities \cite{sat-abs,Kutz}. In terms of this
interpretation, variable $t$ in Eq. (\ref{1D}) designates the
propagation distance, while $x$ is the transverse coordinate [hence
the SP solutions to be produced by Eq. (\ref{1D}) will be spatial solitons \cite%
{we}]. It is also relevant to mention the model combining the standard
fiber-Bragg-grading part and the CQ combination of dissipative terms, in
which temporal DGSs were investigated \cite{NN1}, including interactions
between them \cite{NN2}.

The analysis reported in Ref. \cite{we} revealed, by means of approximate
analytical and direct numerical methods, the existence of three families of
stable DGSs in the first finite bandgap of the respective linear spectrum:
loosely and tightly bound static solitons, and a family of breathers between
them. All the families were found close to the border between the finite
bandgap and Bloch band separating it from the semi-infinite gap, the tightly
and loosely bound DGSs being located in spectral regions where the Bloch
band is, respectively, very narrow or relatively wide. Stable dark solitons
were also found in the model, and the mobility of dark and loosely-bound
bright solitons was demonstrated in it (collisions between mobile solitons
are quasi-elastic).

The objective of the present work is to introduce fundamental and vortical
DGSs in \emph{two dimensions}. To this end, we consider the 2D extension of
Eq. (\ref{1D}),
\begin{eqnarray}
i\frac{\partial \psi }{\partial t} &=&-\frac{1}{2}\nabla ^{2}\psi +|\psi
|^{2}\psi -A\left\{ \cos \left[ 2q_{0}\left( x-\frac{L}{2}\right) \right]
+\cos \left[ 2q_{0}\left( y-\frac{L}{2}\right) \right] \right\} \psi  \notag
\\
&&+i\epsilon (-\gamma _{1}+\gamma _{2}|\psi |^{2}-\gamma _{3}|\psi
|^{4})\psi .  \label{GL}
\end{eqnarray}%
This equation may be interpreted as governing the propagation of
electromagnetic waves in a bulk medium with the self-defocusing nonlinearity
and other ingredients included in Eq. (\ref{1D}), assuming that the
potential periodic in $x$ and $y$ is induced by the 2D transverse grating,
with $t$ again having the meaning of the propagation distance. In Section
II, solutions for fundamental gap solitons are found, in parallel, in a
numerical form [as attractors of Eq. (\ref{GL})], and by means of an
analytical approach, which combines a variational approximation (VA) for the
shape of the solitons and the balance equation for their total power. In
terms of the corresponding linearized equation, the soliton family belongs
to the first finite bandgap. Further, in Section III we report the existence
of stable vortex solitons (with topological charge $1$), built as complexes
of four peaks, with the phase shift of $\pi /2$ between adjacent ones, and
an empty site in the center (\textit{rhombus-shaped} vortices \cite{BBB},
alias \textit{on-site} ones). The vortex solitons are constructed as stable
numerical solutions, and are also obtained by means of the analytical
approximation. Thus, the results reported in this work predict the existence
of stable spatial solitons, both fundamental and vortical ones, in laser
cavities based on self-defocusing bulk media with transverse gratings.

\section{Fundamental solitons}

\begin{figure}[tbp]
\begin{center}
\includegraphics[height=4.5cm]{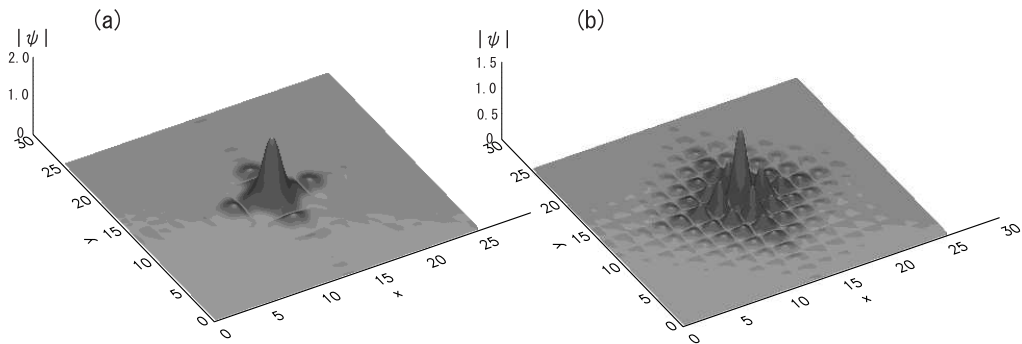}
\end{center}
\caption{Numerically generated profiles of stable fundamental tightly and
loosely bound 2D gap solitons, $|\protect\psi (x,y)|$, found, respectively,
at $q_{0}=1$ (a), and at $q_{0}=1.75$ (b). Other coefficients are $\protect%
\gamma _{1}=0.5,\protect\gamma _{2}=2,\protect\gamma _{3}=1$, $\protect%
\epsilon =0.05$, and $A=2$.}
\label{f1}
\end{figure}

Stable solutions to Eq. (\ref{GL}) in the form of 2D fundamental
(zero-vorticity) solitons can be readily found, in the first finite bandgap,
as \textit{attractors} of the CGL equation, by dint of direct simulations
starting with an appropriate initial configuration. Typical examples of such
\textit{tightly} and \textit{loosely} localized gap solitons are displayed
in Fig. \ref{f1}. The 2D solitons become looser with the decrease of the
lattice period, $\pi /q_{0}$, as the soliton spreads over a larger number of
lattice cells.

The shape of the fundamental solitons is determined by the equation for
complex function $\phi (x,y)$, obtained by the substitution of $\psi \left(
x,y,t\right) =e^{-i\mu t}\phi \left( x,y\right) $ in Eq. (\ref{GL}), where $%
-\mu $ is the soliton's propagation constant, in terms of the optical model.
To approximate the soliton's shape in an analytical form, we start with the
equation for $\phi \left( x,y\right) $ without the dissipative terms ($%
\epsilon =0$). The latter equation can be derived from Lagrangian $\Lambda
=\int \int \mathcal{L}dxdy$, with density
\begin{gather}
\mathcal{L}=\mu |\phi |^{2}-\frac{1}{2}\left( |\nabla \phi |^{2}+|\phi
|^{4}\right)  \notag \\
+A\left\{ \cos \left[ 2q_{0}\left( x-\frac{L}{2}\right) \right] +\cos \left[
2q_{0}\left( y-\frac{L}{2}\right) \right] \right\} |\phi |^{2}.  \label{L}
\end{gather}%
In this approximation, $\phi (x,y)$ may be assumed real, therefore,
following the lines of Refs. \cite{SAM} and \cite{we-recent}, we adopt the
variational ansatz as
\begin{eqnarray}
\phi &=&B\exp \left\{ -\frac{a}{2}\left[ \left( x-\frac{L}{2}\right)
^{2}+\left( y-\frac{L}{2}\right) ^{2}\right] \right\}  \notag \\
&&\times \cos \left[ q\left( x-\frac{L}{2}\right) \right] \cos \left[
q\left( y-\frac{L}{2}\right) \right] ,  \label{GS}
\end{eqnarray}%
where amplitude $B$ and width $a^{-1}$ are free parameters, the norm of
ansatz (\ref{GS}) being
\begin{equation}
N\equiv \int \int \left\vert \phi \left( x,y\right) \right\vert ^{2}dxdy=\pi
B^{2}\left( 4a\right) ^{-1}\left( 1+e^{-q^{2}/a}\right) ^{2}  \label{N}
\end{equation}%
(in terms of the laser-cavity model, $N$ is proportional to the total power
of the light beam).

Ansatz (\ref{GS}) implies that constant $q$ may be different from $q_{0}$ in
Eq. (\ref{GL}). In fact, for the description of tightly-bound (strongly
localized) solitons it will be sufficient to fix $q=q_{0}$, which is not
surprising, as the mismatch between the periodic functions in the ansatz and
the periodicity of the potential, accounted for by $q\neq q_{0}$, is not
crucially important for the well-localized waveforms. On the other hand,
applying the VA to loosely bound (weakly localized) solitons, we will treat $%
q$ not as a variational parameter, but rather as a
``phenomenological" fitting constant. We also tried an extended
version of the VA which subjected $q$ to variation too, but it
yielded essentially less accurate results. This conclusion may be
explained by the difficulty in approximating the complex shape of
loosely-bound solitons by a simple ansatz. Similar situation are
known in other applications of VA to the description of complicated
wave patterns, where some parameters should be still treated as
variational ones, while others are reserved for direct fitting, cf.,
e.g., Ref. \cite{SKA}.

The substitution of ansatz (\ref{GS}) into Lagrangian density (\ref{L}) and
the integration yield the effective Lagrangian, expressed in terms of $N$, $%
a $ and $q$:
\begin{gather}
\Lambda _{\mathrm{eff}}=N\left\{ \mu -\frac{ae^{-q^{2}/a}+a+2q^{2}}{2\left(
1+e^{-q^{2}/a}\right) }\right.  \notag \\
+A\frac{e^{-(q_{0}-q)^{2}/a}+e^{-(q_{0}+q)^{2}/a}+2e^{-q_{0}^{2}/a}}{%
1+e^{-q^{2}/a}}  \notag \\
\left. -\frac{Na}{16\pi }\frac{\left[ 1+e^{q^{2}/\left( 2a\right) }\right]
^{4}\left[ e^{-2q^{2}/a}-2e^{-3q^{2}/\left( 2a\right) }+3e^{-q^{2}/a}\right]
^{2}}{(1+e^{-q^{2}/a})^{4}}\right\} .  \label{Lagr}
\end{gather}%
Variational equation $\partial \Lambda /\partial a=0$ for fixed $q$
determines $a$ as a function of $N$ (as said above, $q$ is not a variational
parameter, hence we do not add equation $\partial \Lambda /\partial q=0$).

At the next stage of the analysis, we restore the dissipative terms in Eq. (%
\ref{GL}) and, treating them as small perturbations, derive a
straightforward evolution equation for the total norm,
\begin{equation}
\frac{dN}{dt}=2\epsilon (-\gamma _{1}N+\gamma _{2}N_{2}-\gamma _{3}N_{3}),
\label{balance}
\end{equation}%
where the following coefficients were calculated as per ansatz (\ref{GS}):%
\begin{equation}
N_{2}=\int \int |\psi |^{4}dxdy=\frac{\pi B^{4}}{128a}e^{-4q^{2}/a}\left[
1+e^{q^{2}/(2a)}\right] ^{4}\left[ 1-2e^{q^{2}/(2a)}+3e^{q^{2}/a}\right]
^{2},  \label{N2}
\end{equation}%
\begin{equation}
N_{3}=\int \int |\psi |^{6}dxdy=\frac{\pi B^{6}}{3072a}\left[
10+e^{-3q^{2}/a}+6e^{-4q^{2}/(3a)}+15e^{-q^{2}/(3a)}\right] ^{2}.  \label{N3}
\end{equation}%
According to Eq. (\ref{balance}), the balance condition for the norm in the
stationary state, $dN/dt=0$, yields relation
\begin{equation}
\gamma _{2}N_{2}=\gamma _{1}N+\gamma _{3}N_{3}.  \label{NNN}
\end{equation}

\begin{figure}[tbp]
\begin{center}
\includegraphics[height=4.cm]{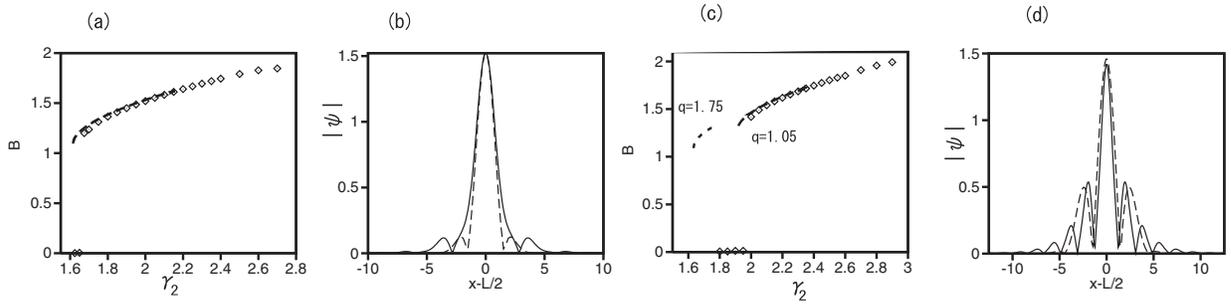}
\end{center}
\caption{(a) Numerically found amplitude $B$ of tightly bound dissipative
gap solitons (diamonds) and the respective analytical prediction (lines),
versus cubic gain $\protect\gamma _{2}$. Other coefficients are $A=2,q_{0}=1,%
\protect\epsilon =0.05,\protect\gamma _{1}=0.5,\protect\gamma _{3}=1$, and $%
q=q_{0}=1$. (b) The comparison of the numerically found and analytically
predicted cross sections of the soliton profiles, $|\protect\psi (x,y=L/2)|$%
, for $q=q_{0}=1$, $\protect\epsilon =0.05,\protect\gamma _{1}=0.5,\protect%
\gamma _{3}=1$, and $\protect\gamma _{2}=2$ (the case of relatively
tightly-bound profiles). (c) Numerically found amplitude $B$ of loosely
bound dissipative gap solitons and the respective analytical prediction,
obtained with fitting parameter $q=q_{0}=1.75$, or $q=0.6q_{0}=1.05$, versus
cubic gain $\protect\gamma _{2}$, for $q_{0}=1.75$. (d) The comparison of
the numerically found and analytically predicted cross sections of loosely
bound soliton profiles for $q_{0}=1.75$, with the choice of $q=1.05$. }
\label{f2}
\end{figure}

Figure \ref{f2}(a) displays the amplitude of the DGS, $B$ for $q_{0}=1$,
versus cubic gain $\gamma _{2}$, as predicted by the analytical
approximation with $q=q_{0}$, i.e., obtained from a numerical solution of
equations (\ref{N}) and (\ref{NNN}), along with the same dependence
generated by direct simulations. It is seen that the analytical
approximation provides good overall accuracy for the solitons which feature
a (relatively) tightly bound shape (therefore, it is sufficient to set $%
q=q_{0}$, i.e., one does not need an extra fitting parameter in this case),
although the approximation predicts the existence of the solitons at $\gamma
_{2}\leq 2.15$, while the direct simulations yield stable DGSs at $\gamma
_{2}\leq 2.7$. For $\gamma _{2}>2.7$, the simulations produce delocalized
solutions, as the loss terms cannot compensate the gain in that region. The
trend to underestimating the existence area for the DGS by the analytical
approximation is generic, persisting throughout the parameter space.

For the same case of $q=q_{0}=1,$ Fig. \ref{f2}(b) displays comparison of
the soliton's cross-section profile, $|\psi (x,y=L/2)|$, and the respective
analytical approximation. For given values of the parameters, the latter one
is $\left\vert \psi (x,y=L/2)\right\vert =$ $1.53e^{-0.413x^{2}}\cos (x-L/2)$%
, see Eq. (\ref{GS}). It is seen that the analytically predicted and
numerically found profiles overlap well near the soliton's center, but
points at which $\phi (x)$ crosses zero are shifted.

Figure \ref{f2}(c) displays the amplitude of \emph{loosely bound solitons}
versus $\gamma _{2}$, for $q_{0}=1.75$. The plot shows both the
straightforward prediction of the analytical approximation, obtained with $%
q=q_{0}=1.75$, and its ``phenomenologically" adjusted modification, provided
by choosing $q=0.6q_{0}=1.05$. In the direct numerical simulations, the
respective DGSs exist for $1.95<\gamma _{2}<2.95$, while the analytical
solution with $q=q_{0}=1.75$ exists in the region of $1.64<\gamma _{2}<1.75$%
, which does not overlap at all with its numerical counterpart. However,
choosing the fitting parameter to be $q=1.05$ allows us to fit the region of
the existence of the analytical solutions to the numerical one fairly well.
Accordingly, the comparison of the numerically found soliton's cross-section
profile, $|\psi (x,y=L/2)|$, and the respective analytical approximation
with $q=1.05$, which is $\left\vert \psi (x,y=L/2)\right\vert =$ $%
1.46e^{-0.151x^{2}}\cos 1.05(x-L/2)$ in the present case, is displayed in
Fig. \ref{f2}(d).

\begin{figure}[tbp]
\begin{center}
\includegraphics[height=4.cm]{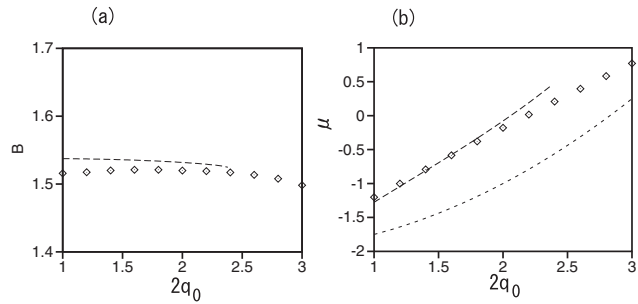}
\end{center}
\caption{Amplitude $B$ (a) and propagation constant $\protect\mu $ (b) of
fundamental solitons versus the grating's wavenumber, $2q_{0}$, for $A=2$
and $\protect\epsilon =0.05,\protect\gamma _{1}=0.5,\protect\gamma _{2}=2$, $%
\protect\gamma _{3}=1$. Diamonds and bold dashed curves show, respectively,
direct numerical results and predictions of the straightforward analytical
approximation, obtained with $q=q_{0}$ (no fitting). The thin dashed curve
in (b) is the lower border of the first finite bandgap [as found with $%
\protect\epsilon =0$, i.e., in the conservative version of Eq. (\protect\ref%
{GL})], to which the soliton family belongs.}
\label{f3}
\end{figure}

Figure \ref{f3} shows another set of global characteristics of the DGS
family, \textit{viz}., the amplitude and propagation constant ($B$ and $\mu $%
) versus the grating's wavenumber, as found in the numerical and analytical
forms. In particular, panel \ref{f3}(b) clearly shows that the entire DGS
family falls into the first finite bandgap of the underlying linear
spectrum. The figure also demonstrates the same trend as mentioned above,
namely, that the analytical approximation, while predicting generally
correct characteristics of the soliton family, underestimates its existence
range: it does not yield solutions for $q_{0}>1.2$, while the direct
simulations converge to DGSs up to $q=1.85$ (numerically found solutions get
delocalized at $q_{0}>1.85$).

\begin{figure}[tbp]
\begin{center}
\includegraphics[height=4.cm]{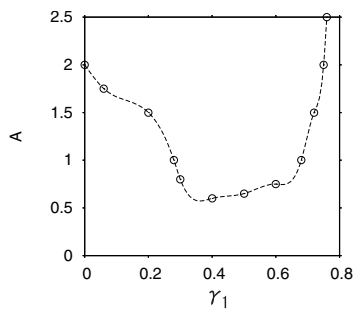}
\end{center}
\caption{Stable fundamental gap solitons exist above the continuous curve in
the $\left( \protect\gamma _{1},A\right) $ plane. Other parameters are $%
\protect\gamma _{2}=2,\protect\gamma _{3}=1,\protect\epsilon =0.05$, and $%
q_{0}=1$.}
\label{f4}
\end{figure}

The existence range for the stable DGSs in the plane of parameters which
control the linear part of the model, \textit{viz}., loss coefficient $%
\gamma _{1}$ and grating strength $A$, as found from the numerical results,
is displayed in Fig. \ref{f4}. To the right of the border shown in Fig. \ref%
{f4} (i.e., if the linear attenuation is too strong), numerical solutions
decay to zero, while to the left of the border (if the attenuation is too
weak) the solution undergoes a delocalization transition. It is relevant to
mention that, unlike DGSs in the 1D counterpart of the present model \cite%
{we}, no 2D soliton was found to be mobile [the application of the \textit{%
kick}, i.e., multiplication by $\exp \left( iKx\right) $, fails to set any
2D soliton in persistent motion].

\section{Solitary vortices}

\begin{figure}[tbp]
\begin{center}
\includegraphics[height=4.5cm]{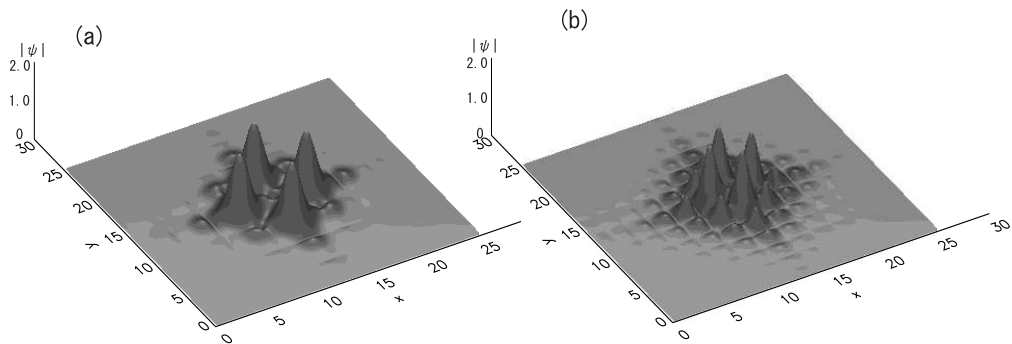}
\end{center}
\caption{Profiles of vortex gap solitons, $|\protect\psi (x,y)|$, produced
by the direct simulations at $q_{0}=1$ (a) and $1.6$ (b) for $\protect\gamma %
_{1}=0.5,\protect\gamma _{2}=2,\protect\gamma _{3}=1$, $\protect\epsilon %
=0.05$, and $A=2$.}
\label{f5}
\end{figure}

We have also found dissipative gap vortex solitons with topological charge $1
$. To generate the vortex soliton, we have performed numerical simulation of
Eq.~(2) starting with the initial configuration
\begin{gather}
\psi \left( x,y\right) =B\left[ \left( x-\frac{L}{2}\right) +i\left( y-\frac{%
L}{2}\right) \right]   \notag \\
\times \exp \left\{ -\left( \frac{a}{2}\right) \left[ \left( x-\frac{L}{2}%
\right) ^{2}+\left( y-\frac{L}{2}\right) ^{2}\right] \right\}   \notag \\
\times \cos \left[ q\left( x-\frac{L}{2}\right) \right] \cos \left[ q\left(
y-\frac{L}{2}\right) \right] ,  \label{gap-vortex}
\end{gather}%
where the pre-exponential multiplier accounts for topological charge $1$ of
the vortex. Typical examples of stable vortices are displayed in Fig.~\ref%
{f5}, at the same values of parameters for which stable fundamental solitons
were shown in Fig. \ref{f1}. Figure \ref{f5}(a) displays a rhombus-type
vortex soliton for $q_{0}=1$, which is composed of four peaks with the phase
shift of $\pi /2$ between adjacent ones. Figure \ref{f5}(b) displays a more
loosely bound vortex soliton, for $q_{0}=1.6$. The solitary vortices feature
a transition from the tightly bound shape to the loose one with the decrease
of the period of the underlying grating, $\pi /q_{0}$. In any case, main
peaks which form vortices are located, approximately, at $\left( x,y\right)
=(L/2+\pi /q_{0},L/2),(L/2-\pi /q_{0},L/2),(L/2,L/2-\pi /q_{0})$ and $%
(L/2,L/2+\pi /q_{0})$. The vortex slowly becomes delocalized at still larger
values of $q_{0}$, for instance at $q_{0}=1.75$, if other parameters are
fixed as in Fig. \ref{f5}

Getting back to the example of a tightly bound vortex obtained at $q_{0}=1$,
which is built of the four peaks which are well localized around four
potential minima, $(x,y)=(L/2+\pi ,L/2),(L/2-\pi ,L/2),(L/2,L/2+\pi )$ and $%
(L/2,L/2-\pi )$, one may approximate each peak by ansatz (\ref{GS}) with $%
q=q_{0}$. Then, the entire vortex may be approximated as
\begin{gather}
\phi \left( x,y\right) =B\cos (x-L/2)\cos (y-L/2)\times   \notag \\
\{\exp [-a/2\{(x-L/2-\pi )^{2}+(y-L/2)^{2}\}]+i\exp
[-a/2\{(x-L/2)^{2}+(y-L/2-\pi )^{2}\}]  \notag \\
-\exp [-a/2\{(x-L/2+\pi )^{2}+(y-L/2)^{2}\}]-i\exp
[-a/2\{(x-L/2)^{2}+(y-L/2+\pi )^{2}]\}  \label{vortex}
\end{gather}%
where $B$ and $a$ are taken as predicted by the analytical approximation for
the fundamental soliton.

\begin{figure}[tbp]
\begin{center}
\includegraphics[height=4.cm]{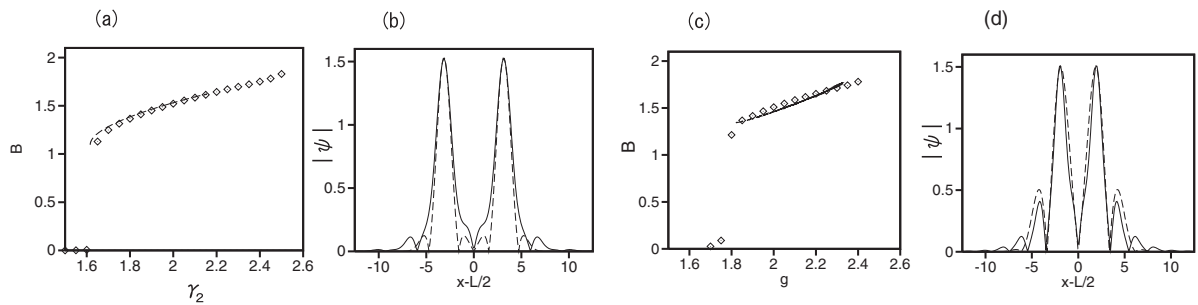}
\end{center}
\caption{(a) The same as in Fig. \protect\ref{f2}(a) (at the same values of
parameters), but for the amplitude of the rhombus-shaped vortex. (b) The
same as in Fig. \protect\ref{f2}(b), but for the cross-section profiles of
the vortex. (c) The numerically found vortex' amplitude for $q_{0}=1.6$, and
the respective analytical approximation (the dashed line) adjusted by
choosing $q=0.7q_{0}=1.12$. (d) The same as in (b), but for $q_{0}=1.6$ and $%
q=1.12$.}
\label{f6}
\end{figure}

Figure \ref{f6}(a) shows the dependence of the vortex' amplitude on the
cubic-gain coefficient, $\gamma _{2}$, as found from the direct simulations
for $q_{0}=1$, i.e., for the family of relatively tightly-bound vortices.
The dashed curve in Fig.~\ref{f6}(a) represents the corresponding analytical
approximation for $B$ in expression (\ref{vortex}), which is found to agree
with the numerical results. Further, Fig. \ref{f6} (b) displays a typical
example of the numerically and analytically found cross-section profiles of
the tightly-bound vortex, which also demonstrates good agreement between
both [the dashed curve is $|\psi (x)|=1.532|\cos (x-L/2)[\exp
\{-0.413(x-L/2-\pi )^{2}\}-\exp \{-0.413(x-L/2+\pi )^{2}\}]|$, as per Eq. (%
\ref{vortex})].

Proceeding to loosely bound vortices, Fig. \ref{f6}(c) shows the dependence
of the vortex' amplitude on $\gamma _{2}$ for $q_{0}=1.6$. The dashed curve
in the same panel represents the respective analytical approximation for $B$%
, adjusted by choosing $q=0.7q_{0}=1.12$ in ansatz (\ref{GS}). Further, Fig. %
\ref{f6}(d) displays a typical example of the numerically and analytically
found cross-section profiles of the vortex, also for $q_{0}=1.6$ and $%
q=0.7q_{0}=1.12$. The dashed (analytical) profile corresponds to $|\psi
(x)|=B\left\vert \cos \left[ q(x-L/2-\pi /q_{0})\right] \exp \left[
-(a/2)(x-L/2-\pi /q_{0})^{2}\right] -\cos \left[ q(x-L/2+\pi /q_{0})\right]
\exp \left[ -(a/2)(x-L/2+\pi /q_{0})^{2}\right] \right\vert $, where $%
B=1.461 $ and $a=0.339$ are values predicted by the analytical approximation
for the fundamental DGS. Good agreement between both profiles is observed.

\section{Conclusion}

In this work, we have introduced a model combining the CGL equation in two
dimensions with the CQ (cubic-quintic) combination of gain and loss terms,
self-defocusing nonlinearity, and the 2D periodic potential. The model gives
rise to stable DGSs (dissipative gap solitons), both fundamental and
vortical ones. The DGS families were found in numerical and approximate
analytical forms in the first finite bandgap of the model's linear spectrum.
The analytical approximation, based on the VA (variational approximation)
for the soliton's shape and balance equation for the total power, yields
results which turn out to be quite accurate in comparison with the numerical
findings (for the fundamental and vortical solitons alike), except for the
fact that the analytical approximation underestimates the region of the
existence of stable DGSs. The model considered in this work can be realized
as a laser cavity in a bulk self-defocusing optical waveguide equipped with
a transverse grating.

The analysis of the model can be developed in other directions. In
particular, it may be interesting to construct solitary vortices with
multiple values of the topological charge (stable 2D gap solitons with
embedded vorticity $2$ were found in the conservative model \cite{vortex-we}%
). A challenging problem is to find DGSs and vortices in higher finite
bandgaps of the underlying linear spectrum.

\end{document}